\documentclass[epj,twocolumn]{webofc}
\usepackage[varg]{txfonts}   % Web of Conferences font

\RequirePackage{pgfcore} % the order of this is important !
\usepackage{lmodern}     % the order of this is important !
\usepackage{graphicx}    % the order of this is important !
\usepackage{tikz}
\usepackage{noam}
\usepackage{lineno}

\usepackage{xcolor,bm,epsfig,wrapfig,Tabbing,axodraw,graphics,graphicx,multirow}
\usepackage[abs]{overpic}
\usepackage{times}

\woctitle{Hadron Collider Physics symposium 2012}
\begin{document}
%\linenumbers
%
\title{Search for heavy resonances, and resonant\\diboson production with the ATLAS detector}
%
% subtitle is optionnal
%
%%%\subtitle{Do you have a subtitle?\\ If so, write it here}

\author{Noam Tal Hod\inst{1}\fnsep\thanks{\email{hod@cern.ch}}
        % etc.
}

\institute{NIKHEF\\ On behalf of the ATLAS collaboration}

\abstract{%
Heavy resonances decaying into a pair of fundamental particles such as $jj$, $\ell^+\ell^-$, $\gamma\gamma$, and $\ell\nu$, are among the most common features to search for phenomena beyond the standard model (SM). Electroweak boson pair production, such as $WW$ or $ZZ$ with subsequent decays to $\ell\nu\ell'\nu'$ and $\ell\ell jj$ respectively, is a powerful test of the spontaneously broken gauge symmetry of the SM and can be also used to search for phenomena beyond the SM. There is a wide spectrum of theoretical models predicting these kinds of resonant signatures. This note covers several searches for these new phenomena conducted within ATLAS in 2011 and 2012 for the LHC 7 and 8~TeV center of mass energies respectively. No significant deviations from the SM have been observed and therefore, limits are set on the characteristic parameters of several new physics models. These benchmark models include new heavy $Z'/W'$ gauge bosons, chiral excitation of the SM weak gauge bosons, $Z^*/W^*$ Randal-Sundrum and ADD gravitons, Composite models for quarks, e.g. $q^*$ with substructure scale $\Lambda$, Quantum black holes, TeV$^{-1}$ Kaluza-Klein excitation of $\gamma/Z$ and more.
}
\maketitle
\section{Introduction}
\label{sec:intro}

This note summarizes several of the ATLAS experiment~\cite{atlas} benchmark analyses as of November 2012, with integrated luminosity ranging from 4.7-5~\ifb for 7~TeV (2011) and 6-13~\ifb for 8~TeV (2012).
In the following, the collection of simple two-body resonances: $\ell^+\ell^-$, $\gamma\gamma$, $jj$, and $\ell\nu$~\cite{dilepton2012,dilepton2011,dilepton2012-CI,diphoton2011-new,dijet2012-2,dijet2011,dijet-old,ellnu2011} and diboson resonances: $WW\to\ell\nu\ell'\nu'$ and $ZZ\to\ell\ell jj$~\cite{ellnuellnu2011,ellelljj2012} will be discussed.
The two types of resonances are covered by a spectrum of models that are described in detail in the ATLAS publications~\cite{dilepton2012,dilepton2011,dilepton2012-CI,diphoton2011-new,dijet2012-2,dijet2011,dijet-old,ellnu2011,ellnuellnu2011,ellelljj2012}\footnote{A comprehensive description of each model can be found in these publications and thereby references.}.
The models discussed by these analyses are:
%\begin{myitemize}
%	\item Heavy gauge bosons, \Zp~\cite{dilepton2012,dilepton2011} and \Wp~\cite{dijet2011,ellnu2011}, from higher symmetry (e.g. \Esix) breaking or from a sequential extension of the SM (denoted as SSM),
%	\item Composite models for quarks, e.g. \qstar, with substructure scale $\Lambda$~\cite{dijet2012-2,dijet2011,dijet-old},
%	\item Randal-Sundrum (RS) gravitons, \Gs\ and \bGs, from warped extra dimensions (EDs)~\cite{dilepton2011,dilepton2012-CI,diphoton2011-new,ellnuellnu2011,ellelljj2012},
%	\item Low-scale string resonances (SR) with large EDs~\cite{dijet2011},
%	\item TeV$^{-1}$ Kaluza-Klein (KK) excitations of \GZsm~\cite{dilepton2011},
%	\item Technicolor (TC) and MWT~\cite{dilepton2011},
%	\item Torsion (TS)~\cite{dilepton2011} and chiral bosons \Zs/\Ws~\cite{dilepton2011,ellnu2011},
%	\item Color octet scalars (s8)~\cite{dijet2011},
%	\item Quantum black holes (QBH)~\cite{dijet2011}, ADD gravitons~\cite{diphoton2011-new}, contact interactions (CI)~\cite{diphoton2011-new,dijet2011}\footnote{The QBH, ADD and CI models predict non resonant signatures but these can be examined by looking on the same experimental signatures as for resonant models.}.
%\end{myitemize}
Heavy gauge bosons, \Zp~\cite{dilepton2012,dilepton2011} and \Wp~\cite{dijet2011,ellnu2011}, from higher symmetry (e.g. \Esix) breaking or from a sequential extension of the SM (denoted as SSM),
Composite models for quarks, e.g. \qstar, with substructure scale $\Lambda$~\cite{dijet2012-2,dijet2011,dijet-old},
Randal-Sundrum (RS) gravitons, \Gs\ and \bGs, from warped extra dimensions (EDs)~\cite{dilepton2011,dilepton2012-CI,diphoton2011-new,ellnuellnu2011,ellelljj2012},
Low-scale string resonances (SR) with large EDs~\cite{dijet2011},
TeV$^{-1}$ Kaluza-Klein (KK) excitations of \GZsm~\cite{dilepton2011},
Technicolor (TC) and minimal walking TC (MWT)~\cite{dilepton2011},
Torsion (TS)~\cite{dilepton2011} and chiral bosons \Zs/\Ws~\cite{dilepton2011,ellnu2011},
Color octet scalars (s8)~\cite{dijet2011},
Quantum black holes (QBH)~\cite{dijet2011}, ADD gravitons~\cite{diphoton2011-new}, contact interactions (CI)~\cite{diphoton2011-new,dijet2011}\footnote{The QBH, ADD and CI models predict non resonant signatures but these can be examined by looking on the same experimental signatures as for resonant models.}.

The connection between the different experimental signatures and the model predictions is given in table~\ref{tbl:models}.
		\begin{table}
		\centering
		\caption{The experimental signatures studied by the ATLAS analyses presented here and the corresponding interpretations, within these analyses. The first three rows correspond to 8~TeV analyses (the rest for 7~TeV).}
		\label{tbl:models}
		{\small
		\begin{tabular}[t]{c|c|c}		
		\hline
		Signature &Models &Publication\\
		\hline		
			$\lPlM$ &\Zp\ &\cite{dilepton2012}\\[3pt]
			$jj$ &\qstar\ &\cite{dijet2012-2}\\[3pt]
			$\ell\ell jj\,(ZZ)$ &\bGs\ &\cite{ellelljj2012}	\\
		\hline
			$\lPlM$ &\Zp,\Gs,TC,KK,TS,\Zs &\cite{dilepton2011}\\[3pt]
			$jj$ &\qstar, QBH, \Wp, CI, Strings &\cite{dijet2011}\\[3pt]
			$\ell\nu$ &\Wp, \Ws\ &\cite{ellnu2011}\\[3pt]
			$\ell\nu\ell'\nu'\,(WW)$ &\Gs, \bGs\ &\cite{ellnuellnu2011}\\[3pt]
			$\gamma\gamma$ &RS, ADD &\cite{diphoton2011-new}\\
		\hline
		\end{tabular}
		}
		\end{table}
This note is divided into two parts, one dealing with simple two-body resonances and the second dealing with diboson resonances.
The separation within each part is signature-wise where for several signatures, there is an overlap with respect to the models interpretation of the limit on the models parameters.

\section{Simple two-body resonances}
\label{sec:twobody}
\subsection{Dilepton}
\label{sec:dilepton}
%Don't forget to give each section, subsection, subsubsection, and
%paragraph a unique label (see Sect.~\ref{sec-1}).
The dilepton signature in this analysis~\cite{dilepton2012,dilepton2011,dilepton2012-CI} is very clean, making possible the identification or exclusion of high mass resonances.
Pairs of high-\pT, isolated $e$ or $\mu$ are selected, where the backgrounds are \GZsm (estimated at NNLO in this analysis), dibosons, \ttbar, multijet and \wjets.
All backgrounds but the \wjets\ are estimated from Monte Carlo (MC) while the \wjets\ (used in the \ee\ channel only) is estimated from the data.
The \GZsm\ contribution can be treated as a part of the signal as it may interfere with the new physics part, as in the KK case, or as a part of the backgrounds if it does not, as in the RS graviton case.
However, in cases where the interference with the new physics amplitude is very small, as in the \Zp\ case, it can be safely neglected and thus the \GZsm contributions can be treated as a part of the backgrounds.
The sum of backgrounds is normalized to the \Zsm\ peak, in the range 70-110~GeV, to cancel out mass-independent uncertainties.
The remaining dominant uncertainties are 20\%  from theory and 21\% from the \ee\ background estimation method.
The data are found to be consistent with the background-only hypothesis with \pvalues 8.6\% (\ee) and 69\% (\mm).
The benchmark model for this analysis is the sequential SM \Zp.
In the 8~TeV analysis~\cite{dilepton2012} it was the only model considered along with several \Esix-inspired \Zps, where the limits on these are always lower than for the \Zpssm.

In figures~\ref{fig:zprime2012} and~\ref{fig:KK}, two of the combined (\ee\ and \mm\ channels) mass limits derived from the dilepton invariant mass distributions are presented.
\begin{figure}
\centering
\includegraphics[width=5cm,clip]{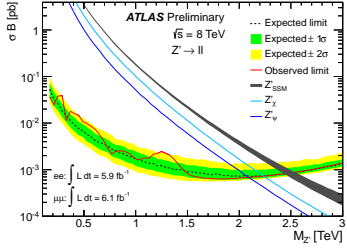}
\caption{95\% CL limits on $\sigmaB(\Zp\to\ell\ell)$ (cross section times branching fraction) vs. \mzp\ for several \Zp\ models~\cite{dilepton2012}.}
\label{fig:zprime2012}
\end{figure}
\begin{figure}
\centering
\includegraphics[width=5cm,clip]{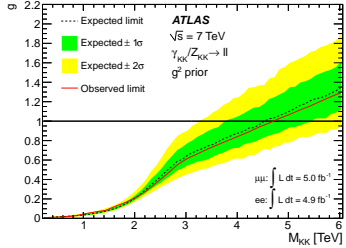}
\caption{95\% CL limit on the KK couplings strength, $g$, vs. \mkk. The strength $g$ is given with respect to the SM partners (\GZsm) couplings~\cite{dilepton2011}.}
%\caption{95\% CL limit on the KK couplings strength, $g$, vs. the KK mass, \mkk.}
\label{fig:KK}
\end{figure}
In figure~\ref{fig:KK}, the limit on \mkk, the mass of the first KK excitation of the \GZsm\ particles, is the highest to date -- almost 1~TeV higher than previous limits obtained from indirect measurements.
It is also the highest direct limit set on a resonance mass in ATLAS.
Table~\ref{tbl:dilepton} lists several representative limits on the parameters of various models, obtained from the two dilepton analyses.
		\begin{table}
		\centering
		\caption{A collection of representative limits on the parameters of various models, as derived from the dilepton mass distributions. The limit in the first row (\Zpssm) was calculated also for 8~TeV~\cite{dilepton2012}, where the number in parentheses corresponds to 7~TeV~\cite{dilepton2011}. The rest of the results are for 7~TeV only~\cite{dilepton2011}.}
		\label{tbl:dilepton}
		{\small
		\begin{tabular}[t]{l|l}
		\hline
			Model &Observed limit [TeV]\\
		\hline
			\Zpssm\ &$M_{\zpssm}\geq 2.49$ $(2.22)$\\[3pt]
		\hline
			\GZkk\ &$\mkk\geq4.71$\\[3pt]
			\Zs\ &$M_{Z^*}\geq 2.20$\\[3pt]
			\Gs\ &$M_{G^*}\geq 2.16$ for $k/\tilde{M}_{\rm Pl} = 0.1$\\[3pt]
			Torsion &$M_{\rm TS}\geq 2.29$ for $\eta_{\rm TS} = 0.2$\\[3pt]
			MWT &$M_A\geq 1.56$ for $\tilde{g} = 2$\\[3pt]
			\hline
			\multirow{3}{*}{Technicolor}
				&$M_{\rho_{\rm T}/\omega_{\rm T}}\geq 0.85$\\[3pt]
				&for $M_{\rho_{\rm T}}-M_{\pi_{\rm T}} = M_W$\\[3pt]
			\hline
			\multirow{3}{*}{Minimal \Zp}
				&$M_{Z'_{\rm min}}\geq 1.11-2.10$\\[3pt]
				&for $\gamma' = 0.2$, $\theta\in [0.\pi]$\\[3pt]
		\hline
		\end{tabular}
		}
		\end{table}
The dilepton signature can be utilized to search also for non-resonant signatures predicted by the set of CI models~\cite{dilepton2012-CI}.

\subsection{Diphoton}
\label{sec:diphoton}
In this analysis~\cite{diphoton2011-new}, the $\gamma\gamma$ results are combined with the \ellell\ results~\cite{dilepton2011} and~\cite{dilepton2012-CI} (see section~\ref{sec:dilepton}) for the applicable models (7~TeV only).
This is another clear signature where one looks on the invariant mass of the two highest-\eT\ ($>$25~GeV) isolated photons in an event.
The benchmark model in this analysis is the RS graviton.
%In this analysis, an ambient energy correction (from low-\pT\ jets) is applied to remove contributions from underlying event and pile-up.
The backgrounds for this analysis are divided into two: (a) {\it irreducible:} SM $\gamma\gamma$ (estimated at NLO from MC), and (b) {\it reducible:} $\gamma j$, $j\gamma$ and $jj$ with one or two jets faking photons (estimated from the data).
The first two options differ by the identity of the leading-\pT object -- the photon or the jet (fake photon).
The sum of backgrounds is normalized in the control region ($142<\mgg<409$~GeV) to cancel out mass-independent uncertainties.
The reducible background is extrapolated to high masses using a smooth function $f(\mgg) = p_1\cdot\mgg^{p_2+p_3\log{\mgg}}$.
The dominant uncertainties in this analysis are: 9\% (photon identification and isolation), 5-15\% (mostly due to parton distribution functions (PDFs)).
The data are found to be consistent with background-only hypothesis with \pvalue 86\%.

Figure~\ref{fig:diphoton} shows the limit on the RS graviton mass for few values of the dimensionless couplings, \kappa/\mpt.
\begin{figure}
\centering
\includegraphics[width=5cm,clip]{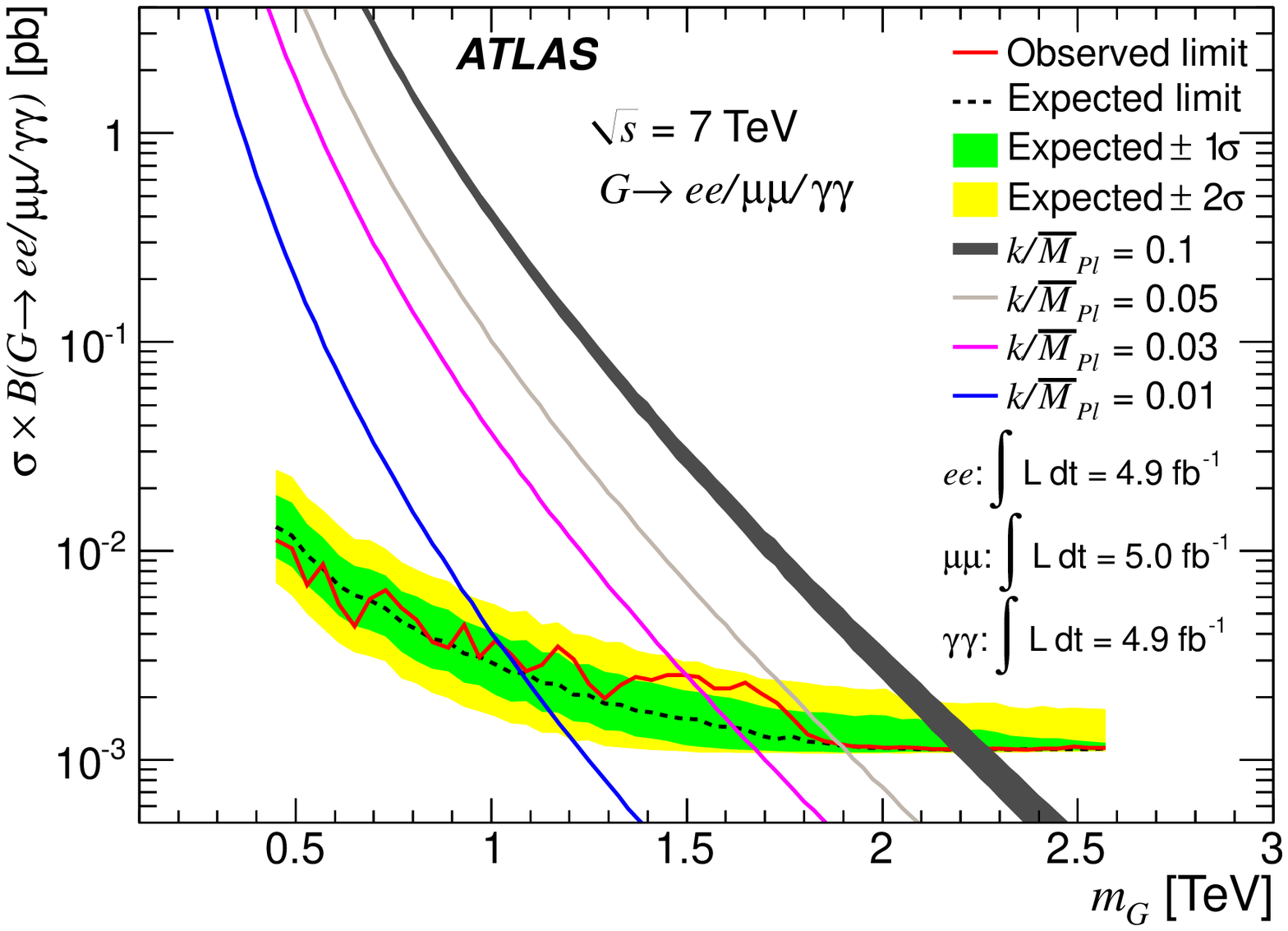}
\caption{95\% CL limits on $\sigmaB(\Gs\to\mm+\ee+\gamma\gamma)$ vs. $M_{\Gs}$ for few values of \kappa/\mpt~\cite{diphoton2011-new}.}
\label{fig:diphoton}
\end{figure}
Combining the result with dileptons yields a limit of 1.03 and 2.23~TeV on the RS graviton mass, for $\kappa/\mpt=0.01$ and 0.1 respectively.
The diphoton channel can be utilized to search also for non-resonant signatures such as those predicted by the ADD model for example.
In that case, the limit on the number of ADD signal events in the search region at $\mgg>1217$~GeV is 7.21 and the lower limits on the ultraviolet cutoff of the KK spectrum, $M_S$, range between 2.79 and 4.18~TeV depending on the number of EDs and the theoretical formalism (see~\cite{dilepton2012-CI}).
%This result is combined with dileptons~\cite{dilepton2012-CI}.

\subsection{Dijet}
\label{sec:dijet}
In this analysis~\cite{dijet2012-2,dijet2011,dijet-old}, the mass and angular distribution of the two most high-\pT jets in an event are used.
The benchmark model used is the excited quark, \qstar.
%The angular distributions can be used here as well, since dominant $t$-channel QCD interactions lead to angular distribution that peak at small scattering angles while the signal is expected to be more isotropic.
The analysis takes advantage of the fact that the dominant $t$-channel QCD interactions lead to angular distribution that peak at small scattering angles while the signal is expected to be more isotropic.
Therefore, the angular distributions are used (in addition to the dijet mass distribution).
%A specific jet energy scale correction is applied to account for the effects on the jet response from additional interactions within the same bunch crossing (``in-time pileup'') and from interactions in bunch crossings preceding or following the one of interest (``out-of-time pileup''). It restores the calorimeter energy scale, on average, to a reference point where pileup is not present.
%Jets are calibrated to the hadronic scale using constants that are functions of the jet \pT\ and pseudorapidity.
%A \pT\ and $\eta$ dependent jet energy scale uncertainty as low as 4\% in the central detector region is assigned to calibrated jets
%For the mass analysis, a data-driven background estimate is used for the \mjj\ spectrum using a smooth function $f(x) = p_1(1-x)^{p_2}x^{p_3+p_4 \ln(x)}$ in $x\equiv \mjj/\sqrt{s}$, whereas for the angular analyses the background estimate relies on QCD MC.
A data driven background estimation leading to the \mjj\ spectrum is parametrized by the function $f(x) = p_1(1-x)^{p_2}x^{p_3+p_4 \ln(x)}$ where, $x\equiv \mjj/\sqrt{s}$, while the background estimate for the angular analyses relies on QCD MC.
The dominant uncertainties for the mass analysis are: 4\% for $\pT^{\rm jet}>1$~TeV from the jet energy scale (JES) and 3.6-3.9\% from the luminosity uncertainties.
The dominant uncertainties for the angular analysis are: $<$8\% due to NLO QCD renormalization and factorization scales, and $<$15\% due to JES.
%Uncertainties (angular): 8\% due to NLO QCD scale variations, and 15\% due to JES
The angular analysis (7~TeV) employs ratio observables called \chi\ and \Fchi\ (see~\cite{dijet2011} for definition) to reduce its sensitivity to systematic uncertainties (JES, PDFs, luminosity) and is more sensitive to non-resonant signals than the mass analysis.
The data are found to be consistent with background-only hypothesis with \pvalue\ 61\%.

Figures~\ref{fig:qstar} and~\ref{fig:gaussian} show the limit on the cross section times acceptance vs. the resonance mass for \qstar\ and for a generic Gaussian model (see~\cite{dijet2011,dijet-old} on how to interpret the Gaussian limits).
%Figure~\ref{fig:qstar} shows the limit on the cross section times acceptance vs. the mass for \qstar.
\begin{figure}
\centering
\includegraphics[width=5cm,clip]{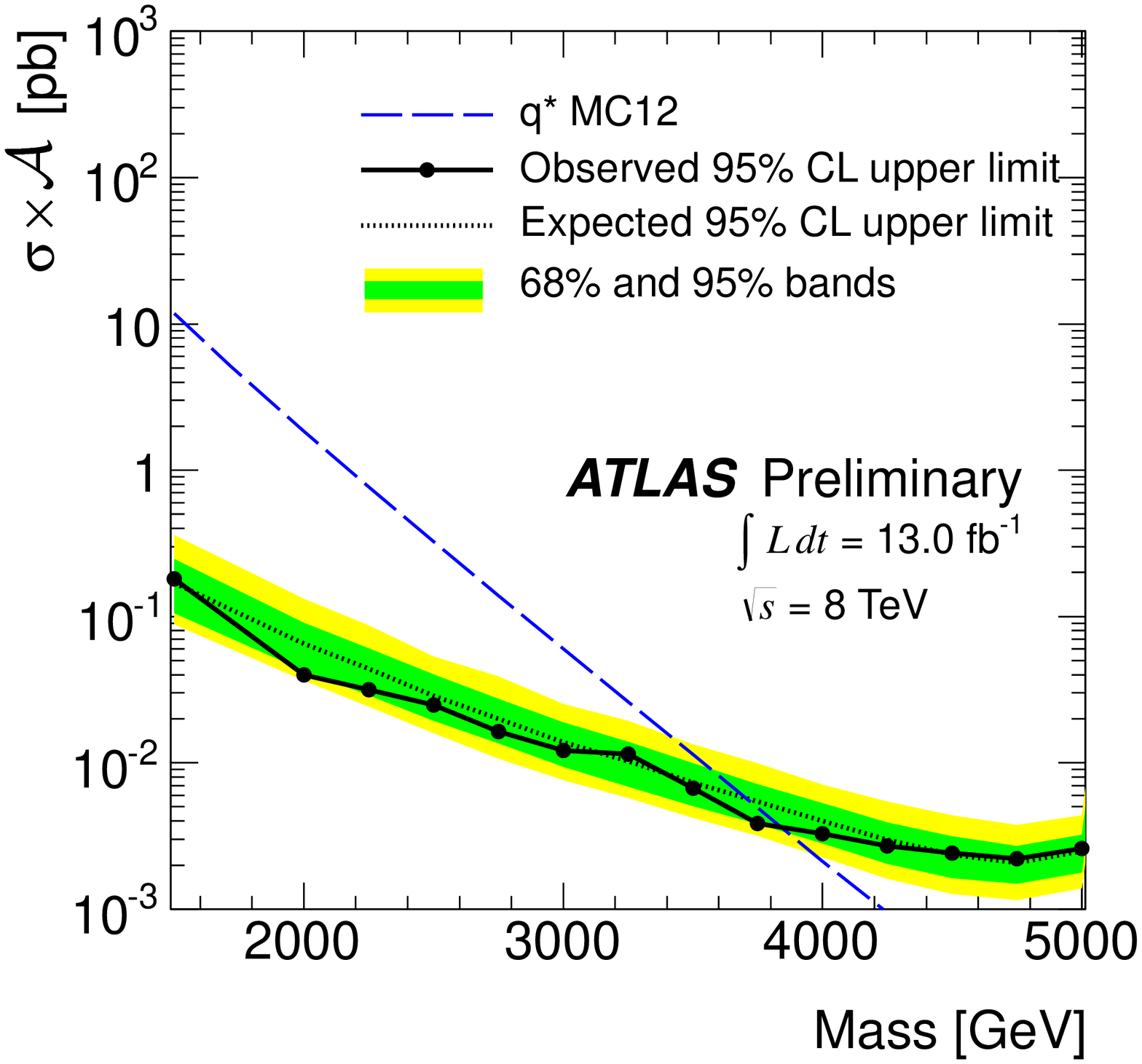}
\caption{95\% CL limits on $\sigma\mathcal{A}(\qstar)$ vs. \mqstar~\cite{dijet2012-2}.}
\label{fig:qstar}
\end{figure}
\begin{figure}
\centering
\includegraphics[width=5cm,clip]{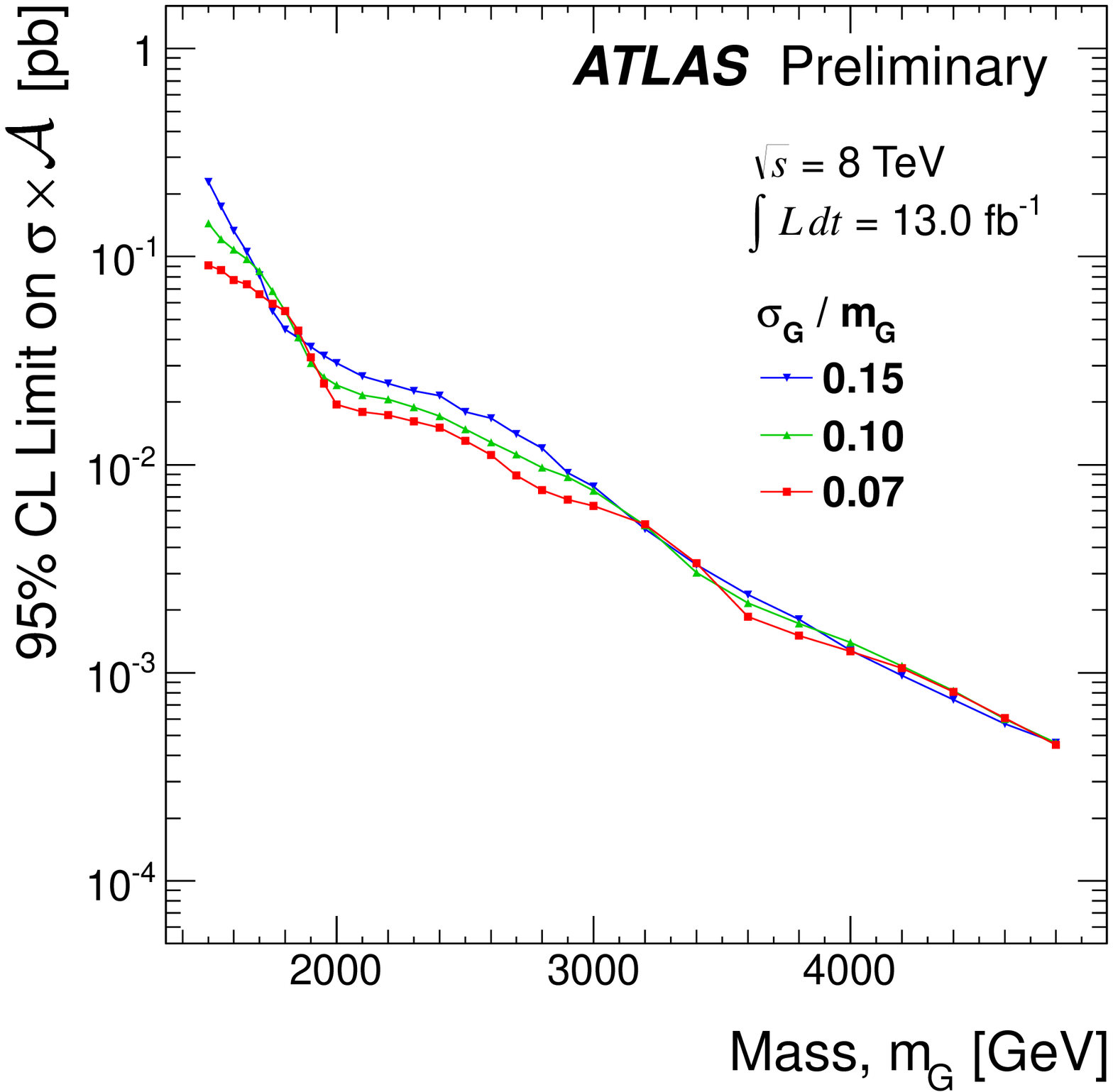}
\caption{95\% CL limits on $\sigma\mathcal{A}({\rm Gaus'})$ vs. $M_{\rm Gaus'}$~\cite{dijet2012-2}.}
\label{fig:gaussian}
\end{figure}
Table~\ref{tbl:dijet} lists several representative limits obtained from both the mass and the angular distribution analyses.	
		\begin{table}
		\centering
		\caption{A collection of representative limits on the parameters of various models derived from the dijet mass and angular distributions. The limit in the first row (\qstar) was calculated also for 8~TeV~\cite{dijet2012-2}, where the number in parentheses corresponds to 7~TeV~\cite{dijet2011}. The rest of the results are for 7~TeV only~\cite{dijet2011}. The last two rows correspond to limits on non-resonant models, obtained from the angular distribution only~\cite{dijet2011}.}
		\label{tbl:dijet}
		{\small
		\begin{tabular}[t]{l|lc}
		\hline
			Model &\multicolumn{2}{c}{Observed limit [TeV]}\\[3pt]
			\,    &from \mjj\ or \chi\ &from \Fchi\\[3pt]
		\hline
			\qstar &$M_{\qstar} \geq 3.84$ (2.83) &--\\[3pt]
		\hline
			Color octet scalar&$M_{\rm s8} \geq 1.86$ &--\\[3pt]
			$W'$ (SM couplings) &$M_{W'}\geq 1.68$ &--\\[3pt]
			String resonances &${\rm SR \,\, scale}\geq 3.61$ &--\\[3pt]
		\hline
			QBH for 6 (2) EDs &$M_D\geq 4.11$ &4.03 (3.71)\\[3pt]
			CI (destructive int') &$\Lambda \geq 7.6$ &$7.6$\\[3pt]
		\hline
		\end{tabular}
		}
		\end{table}
%The limit is also calculated for a generic Gaussian model with different widths.
%See~\cite{dijet2011,dijet-old} on how to interpret the Gaussian limits.
%The acceptance values needed for this interpretation are listed in~\cite{dijet2011}.

\subsection{Charged lepton and a neutrino}
\label{sec:ellnu}
This analysis~\cite{ellnu2011} requires exactly one isolated, high-\pT\ muon or electron with $\pT>$25 and 85~GeV respectively.
A missing transverse energy (\etmis) with same thresholds is also required.
The kinematic variable used to identify a $\Wp/\Ws$ signal is the transverse mass, $m_T = \sqrt{2\pT^{\ell}\etmis(1-\cos\phi_{\ell\nu})}$.
These two signals are distinguishable with respect to both the \mT\ and the angular distributions.
%Different $\mT > \mT^{\rm min}$ threshold is applied per $\Wp/\Ws$ mass and decay channel to discriminate against \wjets\ and QCD backgrounds.
The dominant backgrounds are: SM $W$ bosons (estimated at NNLO), $Z\to\ell\ell$ with one lepton not reconstructed, $\tau$'s from $W/Z$ and diboson production, $\ttbar$ and single-top production and QCD where a hadron decays semileptonically or a jet is misidentified as an electron.
The dominant uncertainties are: 12\% (cross sections) and $<$5\% (experiment).
The data are found to be consistent with the background-only hypothesis.

Figure~\ref{fig:wprime} shows the limit on the cross section times branching fraction vs. $M_{\Wp}$ where the corresponding scenario for the $W^*$ model is comparable.
For the \Wp\ theory curve in figure~\ref{fig:wprime}, no interference with SM $W$ is taken.
This is again a sequential SM scenario.
For the \Ws\ limit, the \Ws\ is taken with $q$ and $g$ coupling strengths normalized to reproduce the \Wp\ width.
The limit on $M_{\Wp}$ ($M_{W^*}$) is 2.55 (2.42)~TeV.
\begin{figure}
\centering
\includegraphics[width=5cm,clip]{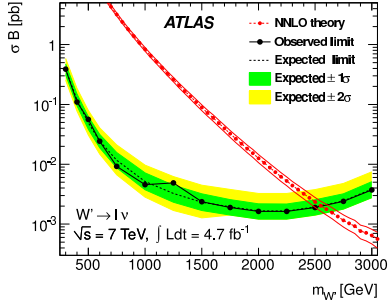}
\caption{95\% CL limits on $\sigmaB(\Wp)$ vs. $M_{\Wp}$ for the combination of the $e$ and $\mu$ channels~\cite{ellnu2011}.}
\label{fig:wprime}
\end{figure}

\section{Diboson resonances}
\label{sec:diboson}
\subsection{WW resonance (\boldmath$\ell\nu\ell'\nu'$)}
\label{sec:ww}
In this analysis~\cite{ellnuellnu2011}, the signature is less pronounced than for the simple two-body resonances because of the subsequent decay of the $W$ bosons produced in the hard interaction.
More over, the presence of two $W$ particles implies a large and non-resolved \etmis. 
In this analysis, one requires exactly two oppositely-charged isolated, high-\pT ($>$25~GeV) leptons, and a large \etmis\ ($>$30, 60 and 65~GeV for $e\mu$, \ee\ and \mm).
In order to reduce $Z$ and top backgrounds, it is required that $\mll>106$~GeV and events with $b$-jets are discarded (the tagging efficiency of $b$-jets is $\epsilon_{b}^{\rm tag}\sim 85\%$).
The $WW$ transverse mass, $m_T^{WW} = \sqrt{(\sum\limits_{i=1}^2{\pT^{\ell_i} + \etmis})^2  -  \sum\limits_{k=x,y}(\sum\limits_{i=1}^2{\p_k^{\ell_i} + E_k^{\rm miss}})^2}$, is used as a discriminant variable.
The RS graviton is again the benchmark model.
However, since this analysis concentrates on the possibility of decay into a pair of heavy $W$ particles, the sensitivity for another theoretical variation of the ordinary RS model is greater.
In the ``bulk'' RS graviton, \bGs\ model, where the ED setup is slightly different than for the ordinary RS model, the graviton has enhanced couplings to the heavier particles, leading to large branching fractions for these states, e.g. $Br(\bGs\to WW)\sim 15\%$.
The dominant backgrounds are: SM $WW$ (estimated at NNLO) and $WZ/ZZ$ with only two reconstructed leptons, $W\gamma$ where the $\gamma$ is reconstructed as a lepton, $\ttbar$ (estimated at NLO, with zero $b$-jets) and single-top (zero $b$-jets), $W/Z$+jets and QCD multi-jet production (both estimated from data).
%The MC predictions for top and \zjets\ events were tested in control regions where the \wjets\ and QCD contributions were estimated from data.
The main uncertainties are: $<$5\% (due to muon resolution correction), 2-9\% (JES), 3.5\% (\etmis\ energy scale), 6-21\% ($\epsilon_{b}^{\rm tag}$ estimation) and 5-10\% (cross sections) and 10-30\% (\wjets\ estimation).
%Further uncertainties: $<$40\% (parton shower model for \ttbar\ events) and $<$10\% ($WW$ modeling)
The data are found to be consistent with background-only hypothesis with \pvalue$>$8\%.

%Figures~\ref{fig:wwgraviton} and~\ref{fig:wwbulkgraviton} show the limit on the cross section times branching fraction vs. the RS graviton and the ``bulk'' RS graviton mass respectively.
%\begin{figure}
%\centering
%\includegraphics[width=5cm,clip]{eps/ellnuellnu/fig_02a}
%\caption{95\% CL limits on $\sigmaB(\Gs\to WW)$ vs. $M_{\Gs}$. }
%\label{fig:wwgraviton}
%\end{figure}
Figure~\ref{fig:wwbulkgraviton} shows the limit on the cross section times branching fraction vs. $M_{\bGs}$.
\begin{figure}
\centering
\includegraphics[width=5cm,clip]{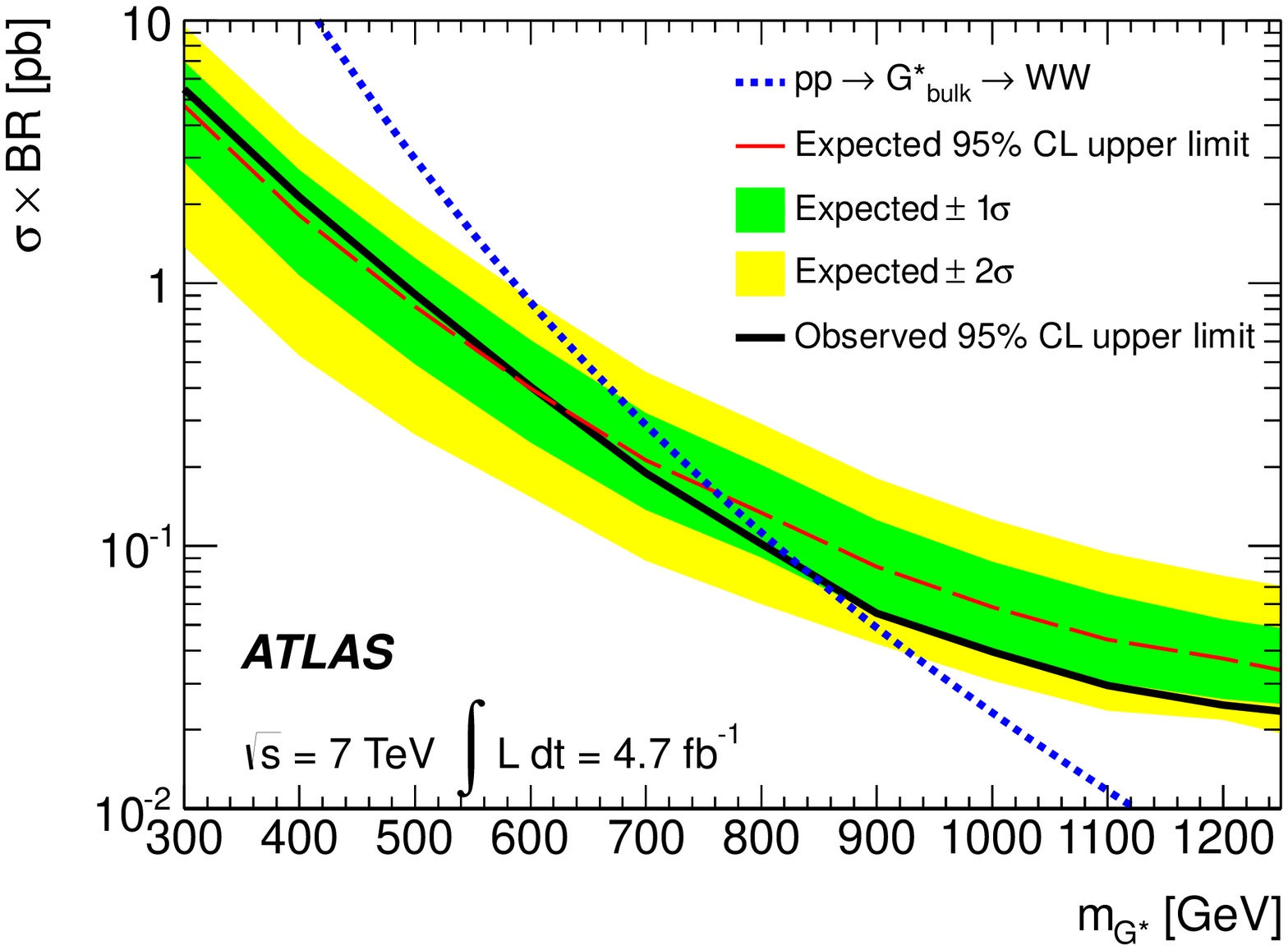}
\caption{95\% CL limits on $\sigmaB(\bGs\to WW)$ vs. $M_{\bGs}$~\cite{ellnuellnu2011}.}
\label{fig:wwbulkgraviton}
\end{figure}
The limits on the ordinary RS and the ``bulk'' RS graviton mass are 1.23 and 0.84~TeV respectively.

\subsection{ZZ resonance (\boldmath$\ell\ell jj$)}
\label{sec:zz}
In this analysis~\cite{ellelljj2012}, the signature is again less pronounced than for the simple two-body resonances because of the subsequent hadronic decay of one of the produced $Z$ bosons.
One requires exactly two isolated, high-\pT, same flavor leptons and two high-\pT\ jets (not within $\Delta R=0.3$ around a lepton).
The presence of leptons reduce the multijet background with respect to the fully-hadronic final state and allow a complete kinematic reconstruction of the intermediate states.
The ``bulk'' RS graviton is the only signal considered in this analysis.
For highly boosted $Z$ bosons, the $\eta-\phi$ distance between the two quarks can be parametrized as $R_{qq} \approx 2M_Z/p_T^Z$.
Therefore, in resonances with masses above $\sim$900~GeV, the two quarks can fall within a $\Delta R = \sqrt{\Delta\phi^2 + \Delta\eta^2}=0.4$ cone, resulting in a single reconstructed massive jet where  $m_{jj}\,\,(m_j)\simeq M_Z$.
As a result, two signal selections must be defined, (a) {\it resolved} and (b) {\it merged}.
The background modeling is tested in these two selections for the \mlljj\ (\mllj) distributions respectively.
Muons are required to be oppositely-charged and the \ellell\ invariant mass is required to be in the range 66$<$\mll$<$116~GeV in order to ensure a $Z$ origin.
The dominant backgrounds are: SM \zjets\ (estimated at NLO), $t\bar t$, SM diboson production, \wjets\ and QCD (estimated from data).
The final background is estimated by fitting the \mlljj\ (\mllj) distributions in the data to a smooth function $f(x) = p_1(1-x)^{p_2}x^{-p_3-p_4 \ln(x)}$ where $x\equiv \mlljj/\sqrt{s}$ or $x\equiv \mllj/\sqrt{s}$ for the two selections respectively.
The main uncertainties are: 5\% for \mlljj$<$800~GeV and 10-40\% for \mllj\ (due to the background fit), 11-15\% (overall uncertainty on the signal acceptance due to jet mass scale, luminosity, JES and ISR/FSR modeling).
The data are found to be consistent with background-only hypothesis.

Figure~\ref{fig:zzbulkgraviton} shows the limit on the cross section times branching fraction vs. $M_{\bGs}$ where the two signal selections are combined by showing only limits for one of them in the range where the expected limit is better.
\begin{figure}
\centering
\includegraphics[width=5cm,clip]{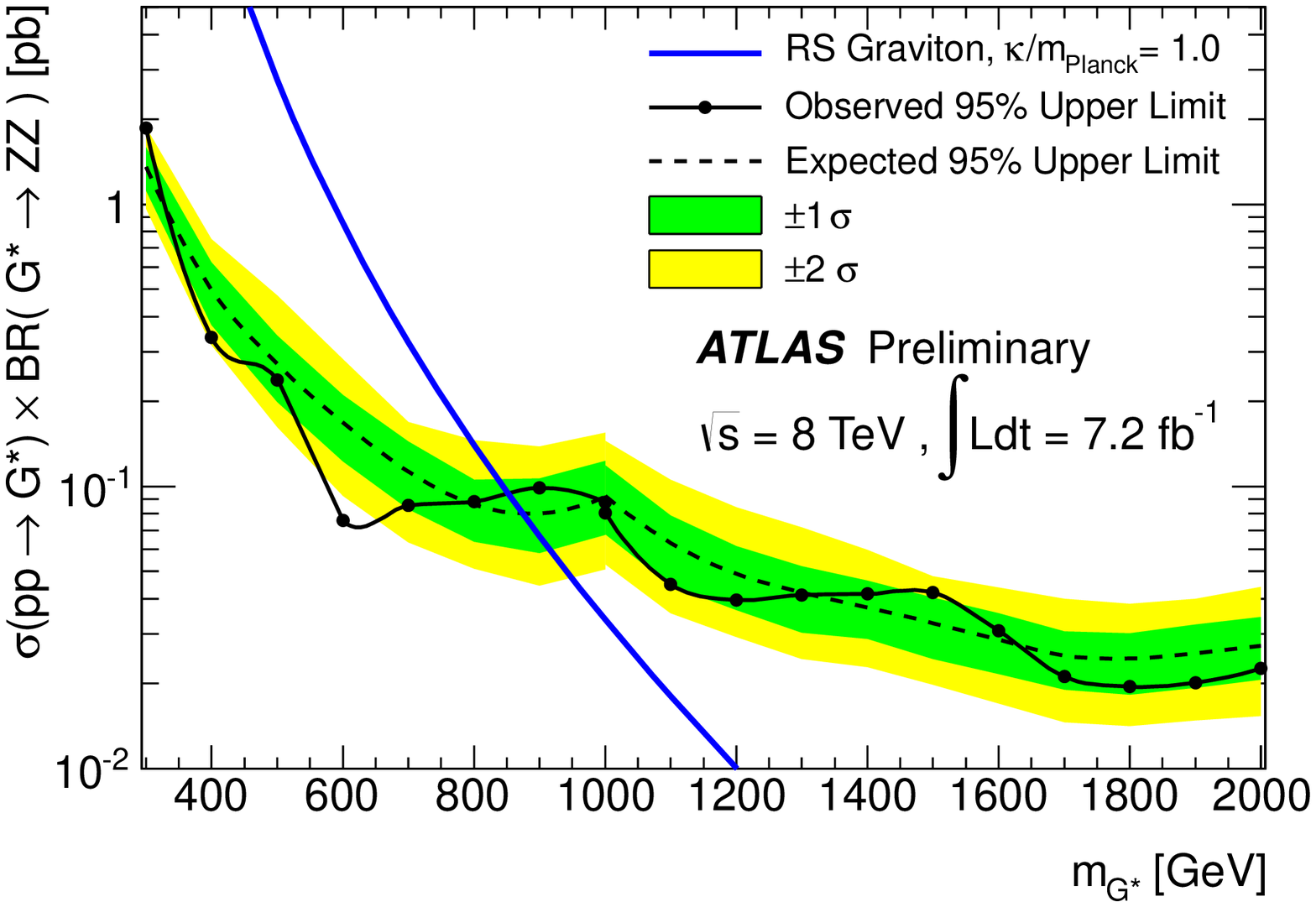}
\caption{95\% CL limits on $\sigmaB(\bGs\to ZZ)$ vs. $M_{\bGs}$~\cite{ellelljj2012}.}
\label{fig:zzbulkgraviton}
\end{figure}
The limit on the ``bulk'' RS graviton mass is 0.84~TeV.

\section{Summary}
\label{sec:summary}
This note covers the most recent ATLAS searches for heavy resonances in 8 different analyses and for 6 different signatures.
The most massive observed event in the data, among these analyses, has a mass of 4.69~TeV (dijet).
No significant deviation from the SM expectations is found in any of these signatures and a set of cutting-edge limits on the parameters of more than 10 models is obtained with a mass reach as high as 4.71~TeV (on \mkk~\cite{dilepton2011}). %and 7.6~TeV on $\Lambda$~\cite{dijet2011} (for the non-resonant CI model).

\end{document}

% end of file template.tex